\documentclass[10pt,
preprintnumbers,amsmath,amssymb,
aps,prd,nofootinbib,eqsecnum,a4paper]{revtex4}
\usepackage{graphicx,epsf}
\usepackage{xcolor}
\def\be{\begin{equation}}
\def\ee{\end{equation}}
\def\bea{\begin{eqnarray}}
\def\eea{\end{eqnarray}}
\def\p{\partial}

\def\cs2{c_{\rm{s}}^2}
\def\wt{\widetilde}
\def\frw{{\rm{FRW}}}
\def\SMTP{{\zeta}_{\rm{SMTP}}}
\def\DOTSMTP{{\dot{\zeta}}_{\rm{SMTP}}}

\newcommand{\sfx}{X}
\newcommand{\sfy}{Y}
\newcommand{\dpn}{\delta P_{\mathrm{nad}}}
\newcommand\eq[1]{Eq.~(\ref{#1})}
\def\beal{\begin{align}}
\def\eeal{\end{align}}
\def\p{\partial}

\begin{document}

\title{Conserved Quantities in Lema{\^\i}tre-Tolman-Bondi Cosmology}
\author{Alexander Leithes and Karim A.~Malik}

\affiliation{Astronomy Unit, School of Physics and Astronomy,
Queen Mary University of London,Mile End Road, London, E1 4NS, UK}

\date{\today}

\begin{abstract}
We study linear perturbations to a Lema{\^\i}tre-Tolman-Bondi (LTB)
background spacetime. Studying the transformation behaviour of the
perturbations under gauge transformations, we construct gauge
invariant quantities. We show, using the perturbed energy conservation
equation, that there are conserved quantities in LTB, in particular a spatial metric trace perturbation, $\SMTP$, which is conserved on all scales. We then briefly
extend our discussion to the Lema{\^\i}tre spacetime, and construct
gauge-invariant perturbations in this extension of LTB spacetime.
\end{abstract}

\maketitle

\section{Introduction}
\label{Introduction}

Conserved quantities are useful tools with a wide range of
applications in cosmology. In particular, they allow us to relate
early and late times in a cosmological model, without explicitly
having to solve the evolution equations, either exactly or taking
advantage of some limiting behaviour. These quantities have been
studied extensively within the context of cosmological perturbation
theory, and usually applied to a Friedmann-Robertson-Walker(FRW)
background spacetime.

Using metric based cosmological perturbation theory
\cite{Bardeen80,KS}, we can readily construct gauge-invariant
quantities which are also conserved, that is constant in time (see
e.g.~Ref.~\cite{Lyth85} for early work on this topic).  
%
%
In a FRW background spacetime, $\zeta$, the curvature perturbation on
uniform density hypersurfaces, is conserved on large scales for
adiabatic fluids.  
%
%
To show that $\zeta$ is conserved and under what conditions, we only
need the conservation of energy \cite{WMLL}. This was first shown to
work for fluids at linear order, but it holds also at second order in
the perturbations, and in the fully non-linear case, usually referred
to as the $\delta N$ formalism \cite{WMLL,MW2003,LMS}\footnote{How the
  gauge invariant curvature perturbation $\zeta$ is constructed is
  briefly discussed in Subsection \ref{FRWST}.}.

Instead of, or in addition to, cosmological perturbation theory, we can
also use other approximation schemes to deal with the non-linearity of
the Einstein equations. In particular gradient expansion schemes have
proven to be useful in the context of conserved quantities, again with the
main focus on FRW spacetimes \cite{Salopek:1990jq, Rigopoulos03, LMS,
  Langlois:2005qp}.
But conserved quantities have also been studied for spacetimes other
than FRW, such as braneworld models (see
e.g.~Ref.~\cite{Bridgman:2001mc}, and anisotropic spacetime
(e.g.~Ref.~\cite{Abolhasani:2013zya}).

The Lema{\^\i}tre-Tolman-Bondi (LTB) spacetime \cite{Bondi} is a more
general solution to Einstein's field equations than the
Friedmann-Robertson-Walker (FRW) model. While LTB is invariant under
rotations, FRW is rotation and translations invariant, and hence has
homogeneous and isotropic, maximally symmetric spatial sections
\cite{Ellis}.

Recent research into LTB cosmology has been motivated by seeking an
alternative explanation for the late time accelerated expansion of the
universe, as indicated by e.g.~SNIa
observations~\cite{Perlmutter:1998np}. Inhomogeneous cosmologies,
including LTB, have been suggested as such an alternative explanation
of these observations (see e.g.~Refs.~\cite{celerier,Marra:2011ct}). Other
observations such as galaxy surveys, large scale structure surveys,
the CMB and indeed any redshift dependent observations
(see for example Refs.~\cite{Moresco:2012jh},~\cite{Anderson:2013zyy},~\cite{Ade:2013ktc})
are usually interpreted assuming a flat FRW cosmology - isotropic and
homogeneous on large scales. In order to test the validity of this
assumption other, inhomogeneous, cosmologies such as LTB should also
be considered. Consequently there is much active research into LTB and
other inhomogeneous spherically symmetric cosmologies, both at
background order and with perturbations (see
e.g.~Refs.~\cite{Iribarrem:2014dta,Lim:2013rra,Biswas,Bolejko1,Bolejko3,CFL,GBH2,MZS10,
ZMS,goodman,Sussman1,Zumalacarregui:2012pq,Clarkson:2012bg,Marra:2011zp} for 
theory and comparison with observation in general, see e.g.~Refs.~\cite{YNS,Alnes,CFZ,ACT,SPT,Marra:2010pg}
for research relating to CMB and see
e.g.~Refs.~\cite{Bull:2011wi,ZS,MZ11,YNS2,SZ80,kSZobs1,kSZobs2,kSZobs3,GBH}
for research more specific to the kinetic Sunyaev-Zeldovich effect, see e.g.~Refs.~\cite{Finelli:2014yha},~\cite{Alonso:2010zv},~\cite{Alonso:2012ds} for structure formation in LTB, including N-body simulations).\\

Gauge-invariant perturbations in general spherically symmetric
spacetimes have been studied already in the 1970s by Gerlach and
Sengupta \cite{Gerlach:1979rw,Gerlach:1980tx}, using a 2+2 split on
the background spacetime. Recent works studying perturbed LTB
spacetimes performs a 1+1+2 split (see e.g.~Refs.~\cite{Tim1},~\cite{February:2012fp},~\cite{February:2013qza}). These
splits allow for a decomposition of the tensorial quantities on the
submanifolds into axial and polar scalars and vectors, similar to the
scalar-vector-tensor decomposition in FRW \cite{Bardeen80,KS}. In this
work we perform a 1+3 split of spacetime, without further decomposing
the spatial submanifold. This prevents us from decomposing tensorial
quantities on the spatial submanifold further into axial and polar scalars and
vectors, but provides us with much simpler expressions, well suited
for the construction of conserved quantities.

We study systematically how to construct gauge-invariant quantities in
perturbed LTB spacetimes. While the existence of conserved quantities in perturbed spacetimes in general is well known, in this paper a conserved gauge invariant quantity, $\SMTP$ is constructed for the first time in LTB (and Lema{\^\i}tre) spacetimes. To this end we derive the transformation
rules for matter and metric variables under small coordinate- or
gauge-transformations. We use these derived gauge-invariant
quantities. We also derive the perturbed energy density evolution
equation, which allows us to derive a very simple evolution equation
for the spatial metric perturbation on uniform density and comoving
hypersurfaces, $\SMTP$. The differences between $\SMTP$ and $\zeta$ in FRW will be discussed in Subsection \ref{LTBst}. The evolution equation for $\SMTP$ and the conditions
under which this variable is conserved will be discussed in Subsection \ref{zeta_evol}.\\

The paper is structured as follows. In the next section we present the
governing equations in covariant form. In Section \ref{ltb_sect} we
review standard Lema{\^\i}tre-Tolman-Bondi cosmology at the background
level.  We then extend the standard results by adding perturbations to
the Lema{\^\i}tre-Tolman-Bondi background. We also look at the
transformation behaviour of matter and metric variables and construct
gauge invariant quantities, including a small review of the procedures
applied to FRW. Finally in this section we look at the evolution of
the spatial metric trace perturbation, $\SMTP$. In Section
\ref{lemaitre} we briefly extend our work to the Lema{\^\i}tre
spacetime in order to allow non-zero background pressure. Finally in
Section \ref{Conclusion} we discuss the implications of our work for
studies of Lema{\^\i}tre-Tolman-Bondi spacetime, and the evolution of
structure therein.

\section{Governing equations}
\label{covariant_sect}

In this section we present the governing equations in covariant form,
that is we do not assume a particular background spacetime, and do
not split quantities into background and perturbations. However, we
assume Einstein's General Relativity (here and throughout this
paper). We also define some covariant quantities that will be useful
in later sections.\\

Einstein's field equations are given by
\begin{equation}
\label{EinsteinGen}
G_{\mu \nu} = 8 \pi G \, T_{\mu \nu}\,,
\end{equation}
where $G_{\mu \nu}$ is the Einstein tensor, $T_{\mu \nu}$ the
energy-momentum tensor, and $G$ is Newton's constant.
The energy-momentum tensor for a perfect fluid is given by,
\begin{equation}
\label{Tmunu_up}
T^{\mu \nu} = (\rho + P) u^{\mu} u^{\nu} + P\ g^{\mu \nu} \,,
\end{equation}
where $g^{\mu \nu}$ is the metric tensor, $\rho$ is the energy density,
$P$ the pressure, and $u^{\mu}$ the 4-velocity of the fluid.
The metric tensor is subject to the constraint,
\begin{equation}
\label{MetricConstraint}
g^{\mu \nu}g_{\nu \gamma} = \delta^{\mu}_{\gamma} ,
\end{equation}
where $\delta^{\mu}_{\gamma}$ is the Kronecker delta.
The 4-velocity is defined by,
\begin{equation}
\label{4vel}
u^{\mu} = \frac{dx^{\mu}}{d\tau}\,,
\end{equation}
where $\tau$ is the proper time along the curves to which $u^\mu$ is tangent, related to the line element $ds$ by
\begin{equation}
\label{Propertime}
ds^2 = - d \tau^2 \,.
\end{equation}
The 4-velocity is subject to the constraint,
\begin{equation}
\label{4velcons}
u^{\mu}u_{\mu} = -1 \,.
\end{equation}
We get energy-momentum conservation from the Einstein equations,
\eq{EinsteinGen}, through the Bianchi identities,
\begin{equation}
\label{Bianchi}
\nabla_{\mu} T^{\mu \nu} = 0 \,.
\end{equation}

The metric tensor allows us to define a unit time-like vector field
orthogonal to constant-time hypersurfaces,
\be
n_\mu \propto \frac{\p t}{\p x^\mu}\,, 
\ee
subject to the constraint
\be
n^\mu n_\mu=-1\,.
\ee
The covariant derivative of any 4-vector can be decomposed as (see for
example \cite{Ellis},~\cite{Wald84}),
\begin{equation}
\label{4vdecomp}
\nabla_\mu n_\nu = -n_\mu u^\alpha \nabla_\alpha n_\nu 
+ \frac{1}{3} \Theta_n {\cal{P}}_{\mu \nu} 
+ \sigma_{\mu \nu} + \omega_{\mu \nu} \,,
\end{equation}
where we use the unit normal vector, $n^\mu$, purely as an example, since \eq{4vdecomp} is true for any 4-vector e.g the 4-velocity, $u^\mu$. Here
$\Theta_n$ is the expansion factor, $\sigma_{\mu \nu}$ the shear
tensor, $\omega_{\mu \nu}$ the vorticity tensor, 
and ${\cal{P}}_{\mu \nu}$ is the spatial projection tensor.

The expansion factor defined with respect to the unit normal vector is, 
\begin{equation}
\label{ExpFacugen}
\Theta_n = \nabla_{\mu} n^{\mu}\,,
\end{equation}

the shear, $\sigma_{\mu \nu}$,
is given by,
\begin{equation}
\label{sheargen}
\sigma_{\mu \nu} 
= \frac{1}{2} {\cal{P}}_{\mu}^{\alpha} {\cal{P}}_{\nu}^{\beta} 
\left( \nabla_{\beta} n_{\alpha} 
+  \nabla_{\alpha} n_{\beta} \right) 
- \frac{1}{3} \Theta_n {\cal{P}}_{\mu \nu} \,,
\end{equation}
where the spatial projection tensor is defined as
\begin{equation}
\label{spatproj}
{\cal{P}}_{\mu \nu} = g_{\mu \nu} + n_{\mu} n_{\nu} \,.
\end{equation}

\section{ Lema{\^\i}tre-Tolman-Bondi spacetime}
\label{ltb_sect}

In this section we first briefly review standard
Lema{\^\i}tre-Tolman-Bondi (LTB) cosmology at the background level. We
then extend the standard results by adding perturbations to the LTB
background. In order to remove any unwanted gauge modes, we study the
transformation behaviour of the perturbations, which then allows us to
construct gauge-invariant quantities, in particular the equivalent to
the curvature perturbation. We show under which conditions this
curvature perturbation is conserved.

Throughout this section we assume zero pressure in the background
spacetime (see Section \ref{lemaitre} for the addition of non-zero background pressure). We do however allow for a
pressure perturbation in the later subsections.

\subsection{Background}
\label{ltb_back}

The LTB metric can be written in various forms \cite{Bondi}, \cite{Bonnor}, \cite{Tim1}. Here we shall
use the following form of the metric \cite{Bondi}, \cite{Ellis},
\begin{equation}
\label{LTBInterval}
ds^2 = -dt^2 + \sfx^2 (r,t) dr^2 
+ \sfy^2 (r,t) \left( d \theta^2 + \sin^2 \theta d \phi^2 \right) ,
\end{equation}
where $\sfx$ and $\sfy$ are scale factors dependent upon both the
radial spatial and time co-ordinates and are not independent, and the
indices $0, 1, 2, 3$ are $t, r, \theta, \phi$ respectively. The scale
factors are related by,
\begin{equation}
\label{ScaleFactorsRelation}
\sfx = \frac{1}{W(r)} \frac{\partial \sfy}{\partial r} ,
\end{equation}
where $W(r)$ is an arbitrary function of $r$, following Bondi
\cite{Bondi}, arising from the Einstein field equations.\\

The 4-velocity in the background is given from its definition,
\eq{4vel}, as
\begin{equation}
\label{4velunpertLTB}
u^{\mu} = [1,0,0,0] \,,
\end{equation}
since we assume we are comoving with respect to the background
coordinates and hence, $dr=d\theta=d\phi=0$, and therefore $d{\tau}^2
= dt^2$ (that is in the local rest frame).

From the definition of the energy-momentum tensor, \eq{Tmunu_up}, we
immediately find that in the absence of pressure the only non-zero
component is, $T^{0 0} = \rho$. For later convenience we define Hubble
parameter equivalents for our two scale factors such that,
\begin{equation}
H_{\sfx} = \frac{\dot{\sfx}}{\sfx} \,, \qquad
H_{\sfy} = \frac{\dot{\sfy}}{\sfy} \,.
\end{equation}
where the ``dot'' denotes the derivative with respect to coordinate time $t$.

The Einstein equations are, from \eq{EinsteinGen}, for the $0-0$
component,
\begin{equation}
\label{HEinstein00}
\frac{1}{\sfy^2} + {H_{\sfy}}^2 + 2\frac{\sfx'\sfy'}{\sfx^3\sfy} + 2H_{\sfx} H_{\sfy} - \left(\frac{\sfy'}{\sfx\sfy}\right)^2 - 2\frac{\sfy''}{\sfx^2\sfy} = 8 \pi G \rho  \,,
\end{equation}
where a prime denotes a derivative with respect to the radial coordinate $r$. 
For the $0-r$ component we find,
\begin{equation}
\label{HEinstein01}
\frac{2}{\sfy} \left(\sfy'H_{\sfx} - \dot{\sfy}' \right) = 0  \,,
\end{equation}
for the $r-r$ component,
\begin{equation}
\label{HEinstein11}
\left(\frac{\sfy'}{\sfx\sfy}\right)^2 - \frac{1}{\sfy^2} - H_{\sfy}^2 - 2\frac{\ddot{\sfy}}{\sfy}  = 0 \,,
\end{equation}
and for ${\theta-\theta}$ and ${\phi-\phi}$ components we get,
\begin{equation}
\label{HEinstein22}
\frac{\sfy''}{\sfx^2\sfy} - \frac{\sfx'\sfy'}{\sfy\sfx^2} 
- \frac{\ddot{\sfy}}{\sfy} - H_{\sfx}H_{\sfy} 
- \frac{\ddot{\sfx}}{\sfx} = 0  \,.
\end{equation}
The other components are identically zero. The energy conservation
equation, obtained from \eq{Bianchi}, is
\begin{equation}
\label{HLTBEngCons}
\dot{\rho} + \rho(H_{\sfx} + 2 H_{\sfy}) = 0\,.
\end{equation}

\subsection{Perturbations}
\label{pert_ltb}

In this section we add perturbations to the LTB background. Unlike
recent works studying perturbed LTB models, e.g.~Refs.~\cite{Tim1}, we
do not decompose the perturbations into polar and axial scalars and vectors, and
multi-poles, which considerably simplifies our governing
equations.

We split quantities into a $t$ and $r$ dependent background part, and
a perturbation depending on all four coordinates. For example we
decompose the energy density $\rho$ as follows,
\be
\label{split_rho}
\rho=\bar\rho(t,r)+\delta\rho(x^\mu)\,,
\ee
where here and in the following a ``bar'' denotes a background
quantity, if there is a possibility for confusion.

We perturb the metric in a similar way as in the flat FRW case, the LTB
metric being very similar to flat FRW in spherical polar
coordinates, save for the two scale factors and the factor of $r$
being absorbed into $\sfy$. 

Hence we split the metric tensor as 
\begin{equation}
\label{metricsplit}
g_{\mu \nu} = {\bar{g}}_{\mu \nu} + \delta g_{\mu \nu} , 
\end{equation}
where ${\bar{g}}_{\mu \nu}$ is given by \eq{LTBInterval}. For the
perturbed part of the metric, $\delta g_{\mu \nu}$, we make the ansatz,
\begin{equation}
\label{LTBMetricperturbations}
\delta g_{\mu \nu}=\begin{pmatrix}
  -2\Phi & \sfx B_r & \sfy B_\theta & \sfy \sin \theta B_\phi \\
  \sfx B_r & 2\sfx^2 C_{rr} & \sfx \sfy C_{r\theta} & \sfx \sfy \sin \theta C_{r\phi} \\
  \sfy B_\theta & \sfx \sfy C_{r\theta} & 2\sfy^2 C_{\theta\theta} & \sfy^2 \sin \theta C_{\theta\phi} \\
  \sfy \sin \theta B_\phi & \sfx \sfy \sin \theta C_{r\phi} & \sfy^2 \sin \theta C_{\theta\phi} & 2\sfy^2 \sin^2 \theta C_{\phi\phi}
\end{pmatrix} \,.
\end{equation}
Here $\Phi$ is the lapse function, and $B_n$, where $n=r,\theta,\phi$,
are the shift functions for each spatial coordinate. Similarly,
$C_{nm}$, where $n,m=r,\theta,\phi$, are the spatial metric
perturbations. As already pointed out, we do not decompose $B_n$ and
$C_{nm}$ further into scalar and vector perturbations (see however
Ref.~\cite{Tim1}).\\

Using the perturbed metric we can construct the perturbed
4-velocities using the definition, \eq{4vel}.
Proper time is to linear order in the perturbations given by,
\begin{equation}
\label{PropCoordPert}
d{\tau} = (1 +  \Phi) dt \,,
\end{equation}
and defining the 3-velocity as $v^i = \frac{d x^i}{d t}$, 
from \eq{4vel} we get the contravariant 4-velocity vector,
\begin{equation}
\label{4velpertLTB}
u^{\mu} = [(1 - \Phi),v^r,v^\theta,v^\phi] .
\end{equation}
By lowering the index using the perturbed metric we obtain the
covariant form,
\begin{equation}
\label{4velpertLTBDown}
u_{\mu} 
= [-(1 + \Phi),
\sfx\left(B_r + {\sfx} v^r\right),\,
\sfy\left( B_\theta + {\sfy} v^\theta \right),\,
\sfy\sin(\theta) \left(B_\phi + {\sfy}\sin(\theta) v^\phi\right)] \,.\\
\end{equation}

Conservation of the energy-momentum tensor, \eq{Bianchi}, allows us
together with its definition, \eq{Tmunu_up}, to derive the perturbed
energy conservation equation, 

\be
\label{PertEconsSphP}
\delta \dot{\rho} + \left(\delta \rho + \delta P\right)
\left(H_X+ 2 H_Y \right) + {\bar{\rho}} ' v^r + {\bar{\rho}}\left(\dot{C}_{rr} + \dot{C}_{\theta\theta} + \dot{C}_{\phi\phi} + {v^r} ' + \partial_\theta v^\theta + \partial_\phi v^\phi + \left[\frac{\sfx '}{\sfx} + 2 \frac{\sfy '}{\sfy} \right] v^r + \cot \theta v^\theta\right) =0 ,
\ee
where we used \eq{split_rho}, and the LTB background requires $\bar P=0$.
The perturbed momentum conservation equations are 
\bea
\label{PertMom1consSphP}
& & \dot{\bar{\rho}} v^r  + \bar{\rho} ({\dot{v}}^r  + \frac{{\dot{B}}_r}{\sfx} + \frac{B_r}{\sfx} H_{\sfx} + (3 H_{\sfx} + 2 H_{\sfy}) v^r) + \frac{1}{\sfx^2} \delta P ' = 0\,,\\
\label{PertMom2consSphP}
& & \dot{\bar{\rho}} v^{\theta} + \bar{\rho} ({\dot{v}}^{\theta}  + \frac{{\dot{B}}_{\theta}}{\sfy} + \frac{B_{\theta}}{\sfy} H_\sfy + (H_{\sfx} + 4 H_{\sfy}) v^\theta) + \frac{1}{\sfy^2} \partial_{\theta} \delta P = 0 \,,\\
\label{PertMom3consSphP}
& & \dot{\bar{\rho}} v^{\phi} + \bar{\rho} \left({\dot{v}}^{\phi}   + \frac{{\dot{B}}_{\phi}}{\sfy \sin \theta} + \frac{{B}_{\phi} H_\sfy}{\sfy \sin \theta} + (H_\sfx + 4 H_\sfy)v^{\phi} \right) + \frac{1}{\sfy^2 \sin^2 \theta} \partial_{\phi} \delta P = 0 \,,
\eea
which we do not use in this work.

\subsection{Gauge Transformation}
\label{GaugeTransBeh}

In order to construct gauge-invariant perturbations, we have to study
the transformation behaviour of our matter and metric variables.  Here
and in the following, we use the Bardeen approach to cosmological
perturbation theory \cite{Bardeen80, KS}. For a general discussion and
references to primary literature, we refer the reader to
Ref.~\cite{MW2008} and references therein.

Using the active point of view, linear order perturbations of a
tensorial quantity $\mathbf{T}$ transform as
\begin{equation}
\label{gauge_gen}
\delta \tilde{\mathbf{T}} = \delta \mathbf{T} 
+ {\pounds}_{\delta x^{\mu}} \bar{\mathbf{T}} \,,
\end{equation}
where the tilde denotes quantities evaluated in the ``new'' coordinate
system.
The old and the
new coordinate systems are related by
\be
\label{coord_sys}
\tilde x^\mu=x^\mu+\delta x^{\mu}\,,
\ee
where $\delta x^{\mu}=[\delta t, \delta x^i]$ is the gauge generator. 
The Lie derivative is denoted by
${\pounds}_{\delta x^{\mu}}$.

\subsubsection{Metric and Matter Quantities}
\label{MetandMat}

From \eq{gauge_gen} and \eq{split_rho} we find that the density
perturbation transforms simply as,
\begin{equation}
\label{denspertgtransLTB}
\delta \tilde{\rho} = \delta \rho +  \dot{{\bar{\rho}}} \delta t
+  {\bar{\rho}}'  \delta r\,,
\end{equation}
since the background energy density depends on $t$ and $r$.
The perturbed spatial part of the 4-velocities, defined in \eq{4velpertLTB}
transform as, 
%
\begin{equation}
\label{vuiTrans}
{\tilde{v}}^i = v^i - \dot{\delta x^i} \,,
\end{equation}
where $i = r, \theta, \phi$.\\

The perturbed metric transforms, using \eq{gauge_gen}, as
\begin{equation}
\label{metricgtransgen2}
{\delta \tilde{g}}_{\mu \nu} = \delta g_{\mu \nu} + \delta x^{\gamma} \partial_{\gamma}  {\bar{g}}_{\mu \nu} +  {\bar{g}}_{\gamma \nu} \partial_{\mu} \delta x^{\gamma} +  {\bar{g}}_{\mu \gamma} \partial_{\nu} \delta x^{\gamma} .
\end{equation}

From the $0-0$ component of \eq{metricgtransgen2} we find that the
lapse function transforms as
\begin{equation}
\label{Phitrans}
\tilde{\Phi} = \Phi - \delta \dot{t} \,.
\end{equation}
For the perturbations on the spatial trace part of the metric we find 
for the $r$ coordinate from \eq{metricgtransgen2},
\begin{equation}
\label{CrrTrans}
{\tilde{C}}_{r r} = C_{r r} + \delta t \frac{\dot{\sfx}}{\sfx} + \delta r \frac{\sfx '}{\sfx} +  \delta r ' \,,
\end{equation}
for the $\theta$ coordinate,
\begin{equation}
\label{CthetheTrans}
{\tilde{C}}_{\theta \theta} = C_{\theta \theta} + \delta t \frac{\dot{\sfy}}{\sfy} + \delta r \frac{\sfy '}{\sfy} + \partial_{\theta} \delta \theta \,,
\end{equation}
and for the $\phi$ coordinate,
\begin{equation}
\label{CphiphiTrans}
{\tilde{C}}_{\phi \phi} = C_{\phi \phi} + \delta t \frac{\dot{\sfy}}{\sfy} + \delta r \frac{\sfy '}{\sfy} + \delta \theta \cot \theta + \partial_{\phi} \delta \phi \,.
\end{equation}

For later convenience
we define a spatial metric perturbation, $\psi$, as,
\begin{equation}
\label{CurvatureLTB1}
3 \psi = \frac{1}{2}\delta g^k_{k} = C_{r r} + C_{\theta \theta} + C_{\phi \phi}\,,
\end{equation}
that is the trace of the spatial metric, in analogy with the curvature
perturbation $\psi_\frw$ in perturbed FRW spacetimes (see Section
\ref{FRWST} below). The relation between $\psi$ here and the curvature
perturbation in perturbed FRW can be most easily seen from the
perturbed expansion scalar, given in \eq{ExpFacSph2} below, which is
very similar to its FRW counterpart (see e.g.~Ref.~\cite{MW2008},
Eq.~(3.19)). The relation is not obvious from calculating the spatial
Ricci scalar for the perturbed LTB spacetime, as can be seen from
\eq{LTBR3EqPert}, given in the appendix.

From the above  $\psi$ transforms as 
\begin{equation}
\label{psiTrans2}
3\tilde{\psi} = 3\psi + \left[\frac{\dot{\sfx}}{\sfx} + 2 \frac{\dot{\sfy}}{\sfy} \right] \delta t + \left[\frac{\sfx '}{\sfx} + 2 \frac{\sfy '}{\sfy} \right] \delta r + \partial_i \delta x^i + \delta \theta \cot \theta \,,
\end{equation}
where $i = r, \theta, \phi$.
In addition, from \eq{metricgtransgen2} the off diagonal spatial metric perturbations transform
as,
\bea
\label{CrtheTrans}
{\tilde{C}}_{r \theta} &=& C_{r \theta} + \frac{\sfy}{\sfx} \delta \theta ' + \frac{\sfx}{\sfy} \partial_{\theta} \delta r \,,\\
\label{CrphiTrans}
{\tilde{C}}_{r \phi} &=& C_{r \phi} + \frac{\sfy \sin \theta}{\sfx} \delta \phi ' + \frac{\sfx}{\sfy \sin \theta} \partial_{\phi} \delta r \,,\\
\label{CthephiTrans}
{\tilde{C}}_{\theta \phi} &=& C_{\theta \phi} + \frac{\sin \theta}{\sfx} \partial_{\theta} \delta \phi + \frac{1}{\sin \theta} \partial_{\phi} \delta \theta \,.
\eea

The mixed temporal-spatial perturbations of the metric, that is the
shift vector, from \eq{metricgtransgen2} transform as
\bea
\label{BrTrans}
{\tilde{B}}_r &=& B_r + \sfx \dot{\delta r} - \frac{\delta t '}{\sfx}\,,\\
\label{BTheTrans}
{\tilde{B}}_{\theta} &=& B_{\theta} + \sfy \dot{\delta \theta} - \frac{\partial_{\theta} \delta t}{\sfy}\,,\\
\label{BPhiTrans}
{\tilde{B}}_{\phi} &=& B_{\phi} + \sfy (\sin \theta) \dot{\delta \phi} - \frac{\partial_{\phi} \delta t}{\sfy (\sin \theta)}\,.
\eea

\subsubsection{Geometric Quantities}
\label{Geom}

The expansion scalar, defined in \eq{ExpFacugen} with $u^\mu$ in place of $n^\mu$, calculated from the
4-velocity, given in \eq{4velpertLTB}, is,
\begin{equation}
\label{ExpFacSph2}
\Theta = \left( H_\sfx + 2 H_\sfy \right) + 3\dot{\psi} + \partial_i v^i  - \left( H_\sfx + 2 H_\sfy \right)\Phi + \left(\frac{\sfx '}{\sfx} + 2 \frac{\sfy '}{\sfy} \right) v^r + \left(\cot \theta \right) v^{\theta} \,,
\end{equation}
where $i = r, \theta, \phi$. Alternatively, the expansion factor
defined with respect to the unit normal vector field defined in
\eq{ExpFacugen}, is given by,
\begin{equation}
\label{ExpFacSphNorm}
\Theta_n = \left( H_\sfx + 2 H_\sfy \right) + 3\dot{\psi} - \left( H_\sfx + 2 H_\sfy \right)\Phi - \frac{B_r '}{\sfx} - \frac{\partial_\theta B_\theta}{\sfy} - \frac{\partial_\phi B_\phi}{\sfy \sin \theta} - \frac{2 B_r \sfy '}{\sfx \sfy} - \frac{B_\theta \cot \theta}{\sfy} \,.
\end{equation}
In order to have the possibility to define later hypersurfaces of
uniform expansion, on which the perturbed expansion is zero, we have
to find the transformation behaviour of the expansion scalar. We find,
that e.g.~$\Theta_n$ transforms as,
\begin{eqnarray}
\label{ExpFacSphNormTrans2}
{\tilde{\Theta}}_n &=& \Theta_n + \left[ {\dot{H}}_\sfx + 2 {\dot{H}}_\sfy \right] \delta t +  \left[ H_\sfx + 2 H_\sfy \right] \dot{\delta t} + \left(\frac{\dot{\sfx} '}{\sfx} - \frac{\dot{\sfx} \sfx '}{{\sfx}^2} + 2 \frac{\dot{\sfy} '}{\sfy} - 2 \frac{\dot{\sfy} \sfy '}{{\sfy}^2} \right) \delta r \\ \nonumber &+& \left[ \frac{1}{\sfx^2} \partial_{r r} + \frac{1}{\sfy^2} \partial_{\theta \theta} + \frac{1}{\sfy^2 \sin^2 \theta} \partial_{\phi \phi} \right] \delta t + \frac{2 \sfy '}{\sfy \sfx^2} \delta t ' + \frac{\cot \theta}{\sfy^2} \partial_\theta \delta t \,.
\end{eqnarray}
We immediately see that the transformation behaviour of $\Theta_n$ is
rather complicated, and we therefore do not use it to specify a gauge.

\subsection{Gauge invariant quantities}
\label{TheCurvePert}

We can now use the results from the previous section, to construct
gauge-invariant quantities. Luckily, we can use the results derived
for the FRW background spacetime as guidance. We follow in particular
Ref.~\cite{WMLL}, which showed that the evolution equation for the
curvature perturbation on uniform density hypersurfaces can be derived
solely from the energy conservation equations (on large scales).\\

\subsubsection{FRW spacetime}
\label{FRWST}

We will first consider the construction of gauge-invariant quantities in perturbed FRW spacetime, which is the homogeneous limit of LTB.
The perturbed FRW metric is \cite{Malik:2004tf}
\be 
\label{ds2}
ds^2=-(1+2\phi)dt^2+2aB_{,i}dt dx^i
+a^2\left[(1-2\psi_\frw)\delta_{ij}+2E_{,ij}\right]dx^idx^j \,, 
\ee
where we have performed a scalar-vector-tensor decomposition, and kept only the scalar part.\ 
\eq{gauge_gen} and \eq{coord_sys} then give \cite{MW2008}
\bea
\label{psi_frw}
\wt{\psi_\frw}&=&{\psi_\frw}+\frac{\dot a}{a}\delta t\,, \\
\wt{\delta\rho_\frw}&=&\delta\rho_\frw+ {\dot {\bar \rho}} \delta t\,, \\
\tilde{E} &=& E + \delta x \,.
\eea
where $a=a(t)$ is the scale factor and $\bar\rho=\bar\rho(t)$ is the
background energy density.
We can now choose a gauge condition, to get rid of the gauge
artefacts, here $\delta t$. To this end, the uniform density gauge can then be specified by the choice $\wt{\delta\rho_\frw}\equiv 0$, which implies
\be
\label{delta_t_frw}
\delta t=-\frac{\delta\rho_\frw}{\dot{\bar{\rho}}}\,.
\ee
Combining \eq{psi_frw} and \eq{delta_t_frw}, we are then led to define
\be
-\zeta \equiv \psi_\frw+\frac{\dot a/a}{\dot\rho}\delta\rho_\frw \,,
\ee
which is gauge-invariant under \eq{gauge_gen}, as can be seen by
direct calculation.

\subsubsection{LTB spacetime}
\label{LTBst}

We can now proceed to construct gauge-invariant quantities in the
perturbed LTB model, taking the FRW case as guidance.  From the
transformation equation of the perturbed spatial metric trace, $\psi$, we see
that here we have to substitute for $\delta t$ and $\delta x^i$, that
is we have to choose temporal \emph{and} spatial hypersurfaces.

From the density perturbation transformation, \eq{denspertgtransLTB}, choosing
uniform density hypersurfaces, $\delta \tilde{\rho}=0$, to
fix the temporal gauge, we get
\be
\label{temporal_ltb}
\delta t\Big|_{\delta \tilde{\rho}=0}= -\frac{1}{\dot{{\bar{\rho}}}}\left[\delta \rho
+  {\bar{\rho}}'  \delta r\right]\,.
\ee
Substituting this into \eq{psiTrans2}, the transformation of the
metric trace, we get
\begin{equation}
\label{zetafull1}
- \SMTP = \psi 
- \frac{1}{3}\left[\frac{\dot{\sfx}}{\sfx} + 2 \frac{\dot{\sfy}}{\sfy} \right] \left(\frac{\delta \rho + \bar{\rho} ' \delta r}{\dot{\bar{\rho}}} \right) 
+ \frac{1}{3}\left\{\left[\frac{\sfx '}{\sfx} + 2 \frac{\sfy '}{\sfy} \right] \delta r + \partial_i \delta x^i + \delta \theta \cot \theta\right\} \,,
\end{equation}
where we chose the sign convention and notation to coincide with the
FRW case.
We can now choose comoving hypersurfaces to fix the remaining spatial
gauge freedom. This gives for the spatial gauge generators from the
transformation of the 3-velocity perturbation, \eq{vuiTrans},
\be
\label{spatial_ltb}
\delta x^i=\int v^i dt\,.
\ee
Substituting the above equations into \eq{zetafull1} we finally get
the gauge-invariant spatial metric trace perturbation on comoving, uniform
density hypersurfaces,
\begin{equation}
\label{zetafull4}
- \SMTP = \psi + \frac{\delta \rho}{3\bar{\rho}} 
+ \frac{1}{3}\left\{
\left(\frac{\sfx '}{\sfx} + 2 \frac{\sfy '}{\sfy}
+ \frac{\bar{\rho} '}{\bar{\rho}} \right) \int v^r dt + \partial_r \int v^r dt + \partial_{\theta} \int v^{\theta} dt + \partial_{\phi} \int v^{\phi} dt + \cot \theta \int v^{\theta} dt\right\} \,,
\end{equation}

i.e. $\SMTP = -\frac{1}{6} \wt{\delta g^k_{k}} \big|_{\wt{\delta \rho} = 0, v = 0}$. We can check by direct calculation, i.e.~by substituting
\eq{psiTrans2}, \eq{denspertgtransLTB}, and \eq{vuiTrans} into
\eq{zetafull4}, that $\SMTP$ is gauge invariant.\\

Instead of using $\delta\rho$ to specify our temporal gauge, we can
just as easily use the spatial metric trace perturbation, that is
define hypersurfaces where $\wt\psi\equiv0$. This gives for $\delta t$
\be
\label{deltatfix}
\delta t= - \frac{1}{H_\sfx + 2 H_\sfy} \left[3 \psi + \left(\frac{\sfx '}{\sfx} + 2 \frac{\sfy '}{\sfy} \right) \delta r + \partial_i \delta x^i + \delta \theta \cot \theta \right].
\ee
This allows us to construct another gauge invariant quantity, the
density perturbation on uniform spatial metric trace perturbation
hypersurfaces, using \eq{denspertgtransLTB}, as
\begin{equation}
\label{GIdenspertZeta}
\delta \tilde{\rho} \Big|_{\psi=0} = \delta \rho + \bar{\rho} \left\{3 \psi +
\left(\frac{\sfx '}{\sfx} + 2 \frac{\sfy '}{\sfy}
+ \frac{\bar{\rho} '}{\bar{\rho}} \right) \int v^r dt + \partial_r \int v^r dt + \partial_{\theta} \int v^{\theta} dt + \partial_{\phi} \int v^{\phi} dt + \cot \theta \int v^{\theta} dt\right\} \,,
\end{equation}
where the spatial gauge generators were eliminated by selecting the
comoving gauge \eq{spatial_ltb} again.
The density perturbation defined in \eq{GIdenspertZeta} can be
written in terms of $\SMTP$, defined in \eq{zetafull4}, simply as
\begin{equation}
\label{GIdenspertZeta2}
\delta \tilde{\rho} \Big|_{\psi=0} = - 3 \bar{\rho} \SMTP\, .
\end{equation}
This expression allows us to relate the density perturbation at different times to the
spatial metric trace perturbation, which, as we shall see in Section
\ref{zeta_evol}, is conserved or constant in time on all scales for barotropic fluids.\\

Alternatively, in both cases above, \eq{zetafull4} and
\eq{GIdenspertZeta}, we could have used the shift functions instead of
the 3-velocities to define the spatial gauge, in analogy with the
Newtonian or longitudinal gauge condition in perturbed FRW. In this
case the spatial gauge generators are
\bea
\label{fixedrshift}
\delta r &=& - \int dt \left[ \frac{\partial_r}{\sfx^2} \left(\frac{\delta \rho}{\dot{\bar{\rho}}} + \frac{B_r}{\sfx} \right) \right] - \int dt \left[ \frac{\partial_r}{\sfx^2} \left(\frac{\delta r \bar{\rho} '}{\dot{{\bar{\rho}}}} \right) \right] \,,\\
%
%
\label{fixedtheshift}
\delta \theta &=& - \int dt \left[ \frac{\partial_\theta}{\sfy^2} \left(\frac{\delta \rho}{\dot{\bar{\rho}}} + \frac{B_\theta}{\sfy} \right) \right] - \int dt \left[ \frac{\partial_\theta}{\sfy^2} \left(\frac{\delta r \bar{\rho} '}{\dot{{\bar{\rho}}}} \right) \right] \,,\\
\label{fixedphishift}
\delta \phi &=& - \int dt \left[ \frac{\partial_\phi}{\sfy^2 \sin^2 \theta} \left(\frac{\delta \rho}{\dot{\bar{\rho}}} + \frac{B_\phi}{\sfy \sin \theta} \right) \right] - \int dt \left[ \frac{\partial_\phi}{\sfy^2 \sin^2 \theta} \left(\frac{\delta r \bar{\rho} '}{\dot{{\bar{\rho}}}} \right) \right] \,.
\eea
Since the expressions are considerably longer than \eq{spatial_ltb} above, we
did not pursue this choice of spatial gauge any further.

Another alternative would be to choose a more geometric definition of
the longitudinal or Newtonian gauge, namely use a zero shear condition
to fix temporal and spatial gauge, again in analogy with FRW, i.e.,
\be
\wt{\delta\sigma_{ij}}=0\,.
\ee
However, again we find that this leads to much more complicated gauge
conditions (since we do not decompose into axial and polar scalar and vector parts), and we
here do not pursue this further. See however appendix \ref{Shear} for
the components of the shear tensor.

For the spatial metric trace perturbation expressed in terms of the perturbations of the Gerlach and Sengupta formalism see appendix \ref{ZetaGS}.

\subsection{Evolution of $\SMTP$}
\label{zeta_evol}

Before we derive the evolution equation for spatial metric trace perturbation
$\SMTP$, we briefly discuss the decomposition of the pressure
perturbation in the LTB setting. We assume that the pressure
$P=P(\rho,S)$, where $\rho$ is the density and $S$ the entropy of the
system. We can then expand the pressure as
\be
\label{deltaP}
\delta P= \cs2\delta\rho+\dpn\,, 
\ee 
where $\dpn$ is the entropy or non-adiabatic pressure perturbation, and
the adiabatic sound speed is define as
\be
\label{cs2def}
\cs2\equiv \left.\frac{\p P}{\p\rho}\right|_S\,,
\ee
for a pedagogical introduction to this topic see
e.g.~Ref.~\cite{Christopherson:2008ry}.
Since in LTB background quantities are $t$ and $r$ dependent, therefore 
allowing for now $P \equiv P(t,r)$, we find that
\be 
\label{cs2_tandr}
\cs2 = \frac{\dot{\bar{P}} + \bar{P}' v^r}{\dot{\bar{\rho}} +
  \bar{\rho}' v^r} \,. 
\ee

However, since in LTB $\bar P=0$, we have that on uniform density
hypersurfaces $\delta P=\dpn$.

The evolution equation for spatial metric trace perturbation on uniform
density and comoving hypersurfaces, $\SMTP$, using the time derivative
of \eq{zetafull4}, \eq{PertEconsSphP} and background conservation
equation, \eq{HLTBEngCons}, is
\begin{equation}
\label{zetaevol5}
 \DOTSMTP =  \frac{H_\sfx + 2 H_\sfy}{3\bar{\rho}} \dpn\, .
\end{equation}
This result is valid on all scales.
We see that $\SMTP$ is conserved for $\dpn=0$, e.g.~for barotropic
fluids. While this result is similar to the FRW case \cite{WMLL},
 we do not have to assume the large scale limit here, which is a striking contrast to be discussed in Section \ref{Conclusion}.

\subsection{Spatial Metric Trace Perturbation in FRW}
\label{CompareFRWSMTP}

In this subsection we will now compare the behaviour of the $\SMTP$ variable that we defined in LTB with the spatial metric trace perturbation on comoving constant density hypersurfaces in FRW spacetime, including background pressure. From \eq{ds2}, the trace of the perturbed part of the spatial metric can be seen to be given in FRW by
\begin{equation}
\label{SMTPtoPsiFRW}
\frac{1}{2} \delta {g^k_{k}}_\frw =  - 3 \psi_\frw + \nabla^2 E \,.
\end{equation}
This quantity can be seen to transform under \eq{gauge_gen} as
\begin{equation}
\label{SMTPPsiFRWTrans}
\wt{\frac{1}{2}\delta g^k_{k}} = \frac{1}{2}\delta {g^k_{k}} -3 H \delta t + \nabla^2 \delta x  \,.
\end{equation}
The 3-velocity transformation has the same form as in LTB, and is given by \eq{vuiTrans}.
Additionally, the density perturbation evolves as
\be
\label{FRWPertE}
\delta \dot{\rho} + 3 H \left(\delta \rho + \delta P\right) - 3 \left(\bar{\rho} + \bar{P}\right) {\dot{\psi}}_\frw + \left(\bar{\rho} + \bar{P}\right) \frac{\nabla^2}{a^2} \left(a v + a^2 \dot{E} \right)
 =0 \,.
\ee
Taking the time derivative of \eq{SMTPtoPsiFRW} and substituting into \eq{FRWPertE} we then find that the spatial metric trace perturbation on comoving constant density hypersurfaces evolves as
\be
\label{FRWSMTP}
\DOTSMTP  = -\frac{1}{6} \dot{\wt{\delta g^k_{k}}} \big|_{\wt{\delta \rho} = 0, v = 0} =  \frac{H}{(\bar{\rho} + \bar{P})} \dpn\, .
\ee
This equation is again valid on all scales, and can again be seen to demonstrate that the spatial metric trace perturbation on comoving constant density hypersurfaces is conserved for barotropic fluids. It should be noted that in order to relate this spatial metric trace perturbation on comoving constant density hypersurfaces in FRW to observables such as the density perturbation both the density perturbation and 3-velocity need to be specified on flat hypersurfaces. It should also be noted that this quantity is not the same as the curvature perturbation, $\zeta$, from the standard FRW literature. Both \eq{FRWSMTP} and \eq{zetaevol5} differ from the result for the Lema{\^\i}tre spacetime, as shall be seen in Section \ref{lemaitre} below.

\section{The Lema{\^\i}tre spacetime}
\label{lemaitre}

Although the main focus of this paper is on LTB cosmology, we here
briefly also discuss perturbations around a Lema{\^\i}tre background
spacetime. The Lema{\^\i}tre spacetime is a generalisation of LTB,
allowing for non-zero pressure in the background
\cite{Hellaby}. Although no exact solution are known in this case, we
nevertheless think it is interesting to extend the discussion of the
previous sections to this spacetime.

The Lema{\^\i}tre background metric is given by
\begin{equation}
\label{LInterval}
ds^2 = -f^2 dt^2 + \sfx^2 (r,t) dr^2 + \sfy^2 (r,t) \left( d \theta^2 + \sin^2 \theta d \phi^2 \right) \,,
\end{equation}
where $f$ is an additional factor, $f \equiv f(t,r)$. The background
four velocity, from \eq{4vel}, is,
\begin{equation}
\label{4velunpertL}
u^{\mu} = \left[\frac{1}{f},0,0,0\right] ,
\end{equation}

and energy-momentum tensor, from \eq{Tmunu_up}, becomes,
\begin{equation}
\label{LStress}
T^{\mu \nu}=\begin{pmatrix}
  \frac{\rho}{f^2} & 0 & 0 & 0 \\
  0 & \frac{P}{\sfx^2} & 0 & 0 \\
  0 & 0 & \frac{P}{\sfy^2} & 0 \\
  0 & 0 & 0 & \frac{P}{\sfy^2 \sin^2 \theta}
 \end{pmatrix} \,.
\end{equation}
Energy conservation is similar to LTB but with an additional pressure term,
\begin{equation}
\label{LEngCons}
\dot{\rho} + (\rho + P)(H_{\sfx} + 2 H_{\sfy}) = 0 \,.
\end{equation}
If we now perturb the metric in a similar way to LTB, \eq{LTBMetricperturbations}, we get,
\begin{equation}
\label{LMetricperturbations}
\delta g_{\mu \nu}=\begin{pmatrix}
  -2 f^2 \Phi & f \sfx B_r & f \sfy B_\theta & f \sfy \sin \theta B_\phi \\
 f \sfx B_r & 2\sfx^2 C_{rr} & \sfx \sfy C_{r\theta} & \sfx \sfy \sin \theta C_{r\phi} \\
 f \sfy B_\theta & \sfx \sfy C_{r\theta} & 2\sfy^2 C_{\theta\theta} & \sfy^2 \sin \theta C_{\theta\phi} \\
 f \sfy \sin \theta B_\phi & \sfx \sfy \sin \theta C_{r\phi} & \sfy^2 \sin \theta C_{\theta\phi} & 2\sfy^2 \sin^2 \theta C_{\phi\phi}
\end{pmatrix} \,.
\end{equation}
The perturbed 4-velocity, from \eq{4vel}, is,
\begin{equation}
\label{4velpertL}
u^{\mu} = \frac{1}{f} \left[(1 - \Phi),v^r,v^\theta,v^\phi \right] \,,
\end{equation}

As in the LTB case, we can now study how the perturbations in this
case change under the transformation \eq{coord_sys}. The perturbed
energy density $\delta\rho$, and the 3-velocities, $v^i$, transform as
in the LTB background \eq{denspertgtransLTB} and \eq{vuiTrans}.
The perturbed metric components transform as
\be
\label{LPhitrans}
\tilde{\Phi} = \Phi - \frac{\dot{f}}{f} \delta t  - \frac{f '}{f} \delta r  + \dot{\delta t} \,,
\ee
and
\bea
\label{LBrTrans}
{\tilde{B}}_r &=& B_r + \frac{\sfx}{f} \dot{\delta r} - \frac{f \delta t '}{\sfx}\,,\\
\label{LBTheTrans}
{\tilde{B}}_{\theta} &=& B_{\theta} + \frac{\sfy}{f} \dot{\delta \theta} - \frac{f \partial_{\theta} \delta t}{\sfy}\,,\\
\label{LBPhiTrans}
{\tilde{B}}_{\phi} &=& B_{\phi} + \frac{\sfy (\sin \theta)}{f} \dot{\delta \phi} - \frac{f \partial_{\phi} \delta t}{\sfy (\sin \theta)}\,.
\eea
The transformation behaviour of the perturbed metric components
$C_{ij}$, and hence $\psi$,  are unchanged from
the LTB case, see \eq{CrrTrans} - \eq{CphiphiTrans} and \eq{psiTrans2} - \eq{CthephiTrans}.\\

The perturbed energy conservation equation is,
\begin{eqnarray}
\label{LPertEconsSphP}
\delta \dot{\rho} &+& \left(\delta \rho + \delta P\right)\left(\frac{\dot{\sfx}}{\sfx} + 2 \frac{\dot{\sfy}}{\sfy}\right) + \left({\bar{\rho}} ' + {\bar{P}} '\right)v^r + \frac{f B_r}{\sfx} {\bar{P}}'+ \left(\partial_\theta \frac{B_\theta}{\sfy} + \partial_\phi \frac{B_\phi}{\sfy \sin \theta}\right) f {\bar{P}} \nonumber \\
&+& \left({\bar{\rho}} + {\bar{P}}\right)\left(\dot{\psi} + {v^r} ' + \partial_\theta v^\theta + \partial_\phi v^\phi + \left[\frac{f '}{f} + \frac{\sfx '}{\sfx} + 2 \frac{\sfy '}{\sfy} \right] v^r + \frac{B_r f '}{\sfx} + \cot \theta v^\theta\right) 
 = 0 \,.
\end{eqnarray}

As in the previous section, we can now construct gauge-invariant
quantities.  We choose hypersurfaces of vanishing perturbed energy
density to define the temporal gauge, that is,
\begin{equation}
\label{Consdenshyp}
\delta t = \frac{\delta \rho}{\dot{\bar{\rho}}}
+ \frac{\bar{\rho}'}{\dot{\bar{\rho}}} \delta r \,,
\end{equation}
and choose again co-moving gauge, where $v^i = 0$, to get for
the spatial coordinate shifts
\begin{equation}
\label{LvuiTransComoving}
\delta x^i = \int v^i dt \,.
\end{equation}
Then using the transformation for perturbed metric trace $\psi$, given
above in \eq{CurvatureLTB1}, we can construct the gauge-invariant
spatial metric trace perturbation on uniform density and comoving hypersurfaces,
\begin{equation}
\label{LPresszetafull4}
- \SMTP = \psi + \frac{\delta \rho}{3(\bar{\rho} + \bar{P})} + \frac{1}{3}\left\{\left(\frac{\sfx '}{\sfx} + 2 \frac{\sfy '}{\sfy}
+ \frac{\bar{\rho} '}{\bar{\rho} + \bar{P}} \right) \int v^r dt + \partial_r \int v^r dt + \partial_{\theta} \int v^{\theta} dt + \partial_{\phi} \int v^{\phi} dt + \cot \theta \int v^{\theta} dt \right\}\,.
\end{equation}
The evolution equation for $\SMTP$ is then found from
\eq{LPertEconsSphP}, using the decomposition of the pressure
perturbation, \eq{deltaP}, and the definition of the adiabatic sound
speed, \eq{cs2_tandr}, as
%
%
\begin{eqnarray}
\label{AdPresszetaevol1b}
- \DOTSMTP &=& 
 \frac{ \dot{\bar{\rho}}}{\left({\bar{\rho}} + {\bar{P}}\right)^2}\dpn - \frac{\bar{P} '}{\left({\bar{\rho}} + {\bar{P}}\right)}v^r 
- \frac{f B_r}{\sfx \left({\bar{\rho}} + {\bar{P}}\right)} {\bar{P}} ' 
- \left(\partial_\theta \frac{B_\theta}{\sfy} + \partial_\phi \frac{B_\phi}{\sfy \sin \theta}\right) \frac{f \bar{P}}{\left({\bar{\rho}} + {\bar{P}}\right)} \\ \nonumber &+& \left[\partial_t \left(\frac{\sfx '}{\sfx} + \frac{\sfy '}{\sfy} + \frac{\bar{\rho} '}{\bar{\rho} + \bar{P}} \right) \right] \int v^r dt - \frac{f '}{f} v^r  + \frac{B_r f '}{\sfx} \,.
\end{eqnarray}

By transforming the coordinates to Cartesian using the chain rule and taking the spatial derivatives to be negligible on large scales, \eq{AdPresszetaevol1b}, reduces to,
\begin{eqnarray}
\label{AdPresszetaevol3}
\DOTSMTP &=&  \frac{H_\sfx + 2 H_\sfy}{3({\bar{\rho}} + {\bar{P}})}
\, \dpn \,.
\end{eqnarray}
This can be seen to be similar to that for LTB, \eq{zetaevol5}, but as with the standard $\zeta$ in FRW and unlike $\SMTP$ in both LTB and FRW is only valid at large scales.

\section{Discussion and conclusion}
\label{Conclusion}

In this paper we have constructed gauge-invariant quantities in
perturbed LTB spacetime.  In particular we have constructed the
gauge-invariant spatial metric trace perturbation on comoving, uniform density
hypersurfaces, $\SMTP$. We derived the evolution equation for $\SMTP$
and found that it is conserved on all scales for barotropic fluids
(when $\dpn=0$). We found this result for the evolution equation for $\SMTP$ also holds for FRW. This is in contrast to the standard FRW result,
where the similar gauge-invariant quantity, $\zeta$, is only conserved on
large scales. It was also found that the evolution equation for $\SMTP$, in Lema{\^\i}tre spacetime which would be conserved in the case of barotropic fluids is only found in the large scale limit, as with the result for the standard $\zeta$ in FRW.\\

Deriving these results in LTB is more involved than in the FRW case, because
the background is $t$ and $r$ dependent, whereas the FRW background is
homogeneous and isotropic, and hence only $t$ dependent.
Additional complications often arise in LTB because it suggests a 1+1+2
decomposition, and not ``simply'' a 1+3 one, as in FRW. This makes a
multi-pole decomposition much more complicated, and hence we did not
use it here to construct conserved quantities.\\



The main application of the results derived in this paper lies in providing conserved quantities in LTB and Lema{\^\i}tre spacetimes. Conserved quantities have proved to be very useful in the FRW case, and they can be used to directly relate observable quantities, such as the density perturbation, between the beginning of the epoch modelled by the LTB spacetime and late times, without having to solve the field equations. Furthermore, these conserved quantities can be used as a consistency check for numerical simulations, as for example those described in ~Ref.\cite{February:2013qza}. By solving for the density contrast numerically and comparing this result to the one obtained by using $\SMTP$, the accuracy of the code can easily be checked.

\begin{acknowledgments}
The authors are grateful to Tim Clifton and David Mulryne for useful and insightful discussions. AL is funded by a
STFC studentship, and KAM is supported, in part, by STFC grant
ST/J001546/1. The computer algebra package
{\sc{Cadabra}}\cite{Cadabra} was used in the derivation of some of the
equations.
\end{acknowledgments}



\appendix

\section{Additional material for LTB}

In this section of the appendix we present some material that is not
essential to follow the main body of this work. However, since it
might be useful and save time in reproducing or extending some or all
of the calculations, we reproduce it here.

\subsection{Contravariant LTB Metric Perturbations}
\label{Contmetpert}

Using the constraint \eq{MetricConstraint} we get the contravariant perturbed
metric components,
\begin{equation}
\label{LTBMetricperturbationsuppy}
\delta g^{\mu \nu}=\begin{pmatrix}
  2\Phi & \frac{B_r}{ \sfx } & \frac{B_\theta}{ \sfy } & \frac{B_\phi}{ \sfy \sin \theta} \\
  \frac{B_r}{ \sfx } & - \frac{2C_{rr}}{\sfx^2} & - \frac{C_{r\theta}}{ \sfx \sfy } & - \frac{C_{r\phi}}{ \sfx \sfy \sin \theta}  \\
  \frac{B_\theta}{ \sfy } & - \frac{C_{r\theta}}{ \sfx \sfy } & - \frac{2C_{\theta\theta}}{\sfy^2}  & - \frac{C_{\theta\phi}}{ \sfy^2 \sin \theta}  \\
 \frac{B_\phi}{ \sfy \sin \theta} & - \frac{C_{r\phi}}{ \sfx \sfy \sin \theta} & - \frac{C_{\theta\phi}}{ \sfy^2 \sin \theta} & - \frac{2 C_{\phi\phi}}{\sfy^2 \sin^2 \theta }
\end{pmatrix} .
\end{equation}

\subsection{LTB Shear}
\label{Shear}
The $t-t$ component of the shear is zero. To linear order we find that the $r-r$ component is,
\begin{eqnarray}
\label{shearrr}
\sigma_{r r} &=& -\frac{1}{3} \sfx^2 \bigg( \dot{\psi} - 2 (1 - \Phi)(H_\sfx - H_\sfy) - 4 C_{r r} (H_\sfx - H_\sfy) - 2 \frac{B_r}{\sfx \sfy} \sfy ' - \frac{B_\theta \cot \theta}{\sfy} \\ \nonumber &+& \frac{2}{\sfx} B_r ' - \frac{1}{\sfy} \partial_\theta B_\theta - \frac{1}{\sfy \sin \theta} \partial_\phi B_\phi - 3 \dot{C_{r r}} \bigg) ,
\end{eqnarray}
the $\theta-\theta$ component,
\begin{eqnarray}
\label{shearthethe}
\sigma_{\theta \theta} &=& -\frac{1}{3} \sfy^2 \bigg( \dot{\psi} + (1 - \Phi)(H_\sfx - H_\sfy) + 2 C_{\theta \theta} (H_\sfx - H_\sfy) + \frac{B_r \sfy '}{\sfx \sfy} - \frac{B_\theta}{\sfy} \cot \theta \\ \nonumber &-& \frac{B_r '}{\sfx} + \frac{2}{\sfy} \partial_{\theta} B_{\theta} - \frac{\partial_\phi B_\phi}{\sfy \sin \theta} - 3 {\dot{C}}_{\theta \theta} \bigg) ,
\end{eqnarray}
and the $\phi-\phi$ component,
\begin{eqnarray}
\label{shearphiphi}
\sigma_{\phi \phi} &=& -\frac{1}{3} \sfy^2 \sin^2 \theta \bigg( \dot{\psi} +  \left( 1 - \Phi \right) \left( H_\sfx - H_\sfy \right) + 2 C_{\phi \phi} \left( H_{\sfx} - H_{\sfy} \right) + \frac{B_r \sfy '}{\sfx \sfy} + 2 \frac{B_\theta}{\sfy} \cot \theta \\ \nonumber &-& \frac{ B_r '}{\sfx} - \frac{\partial_\theta B_\theta}{\sfy} - \frac{\partial_\phi B_\phi}{\sfy \sin \theta} - 3 {\dot{C}}_{\phi \phi} \bigg) .
\end{eqnarray}
We also need the off-diagonal components. For the mixed temporal-spatial components we get,
\begin{equation}
\label{shearti}
\sigma_{t r} = \frac{2 B_r \sfx}{3} \left(H_{\sfx} - H_{\sfy} \right) ,
\qquad
\sigma_{t \theta} = - \frac{B_\theta \sfy}{3} \left(H_{\sfx} - H_{\sfy} \right) ,
\qquad
\sigma_{t \phi} = - \frac{B_\phi \sfy \sin \theta}{3} \left(H_{\sfx} - H_{\sfy} \right) .
\end{equation}

For the mixed spatial components we get,
\bea
\label{shearij}
\sigma_{r \theta} &=& \frac{1}{3} C_{r \theta} \sfx \sfy \left( H_{\sfx} - H_{\sfy} \right) + \dot{C_{r \theta}} \sfx \sfy - \frac{1}{2} \sfy B_{\theta} ' + \frac{1}{2} B_{\theta} \sfy ' - \frac{1}{2} \sfx \partial_{\theta} B_r .
\\
\sigma_{\theta \phi} &=& - \frac{1}{3} C_{\theta \phi} \sfy^2 \sin \theta \left( H_{\sfx} - H_{\sfy} \right) + \frac{1}{2} \dot{C_{\theta \phi}} \sfy^2 \sin \theta - \frac{1}{2} \sfy \sin \theta \partial_\theta B_{\phi} + \frac{1}{2} \sfy \cos \theta B_{\phi} .
\\
\sigma_{r \phi} &=& \frac{1}{6} C_{r \phi} \sfx \sfy \sin \theta \left( H_{\sfx} - H_{\sfy} \right) + \frac{1}{2} \dot{C_{r \phi}} \sfx \sfy \sin \theta - \frac{1}{2} \sfy \sin \theta B_{\phi} ' + \frac{1}{2} \sin \theta B_{\phi} \sfy ' .
\eea

\subsection{The LTB Ricci 3-scalar}
\label{LTBR3}
The Ricci scalar on the spatial 3-hypersurfaces is given, in the background, as,
\begin{eqnarray}
\label{LTBR3EqBack}
{\bar{R}}^{(\mathrm{3})} &=& \frac{4 \sfx ' \sfy '}{\sfx^3 \sfy} - \frac{2  \sfy '^2}{\sfx^2 \sfy^2} - \frac{4 \sfy ''}{\sfx^2 \sfy} + \frac{2}{\sfy^2} ,
\end{eqnarray}
and the perturbed Ricci scalar is given by,
\begin{eqnarray}
\label{LTBR3EqPert}
\delta R^{(\mathrm{3})} &=& \frac{4 C_{r r} }{\sfx^2 \sfy} \left(\frac{\sfy '^2}{\sfy} + 2 \sfy '' - \frac{2 \sfx ' \sfy '}{\sfx} \right) - \frac{2}{\sfy^2} (2 C_{\theta \theta}) - \frac{2}{\sfx^2}(C_{\theta \theta} '' + C_{rr} '') + \frac{2 C_{r \theta} \cot \theta}{\sfx^2 \sfy} (\sfx ' - \sfy ') \\ \nonumber &+& \frac{2 \sfx ' (\partial_\theta C_{r \phi})}{\sfx^2 \sfy} + \frac{2 \sfx ' C_{\theta \theta} '}{\sfx^3} + \frac{4 \sfy '  C_{r r} '}{\sfx^2 \sfy} + \frac{2 \sfx ' C_{\phi \phi} '}{\sfx^3} + \frac{2 \cot \theta (\partial_\theta C_{\theta \theta})}{\sfy^2} \\ \nonumber &-& \frac{6 \sfy ' ( C_{\theta \theta} ' + C_{\phi \phi} ')}{\sfx^2 \sfy} - \frac{4 \cot \theta (\partial_\theta C_{\phi \phi})}{\sfy^2} - \frac{2 (\partial_{\theta \theta} C_{\phi \phi})}{\sfy^2} + \frac{2 (\partial_{\theta} C_{r \theta} ')}{\sfx \sfy} + \frac{2 (\partial_{\phi} C_{r \phi} ')}{\sfx \sfy \sin \theta} + \frac{2 (\partial_{\theta \phi} C_{\theta \phi})}{\sfy^2 \sin \theta} .
\end{eqnarray}

\section{The Spatial Metric Trace Perturbation in 2+2 Spherical Harmonic Formalism}
\label{ZetaGS}

\subsection{Background}
\label{GSBack}

The background LTB metric in the Clarkson, Clifton and February formalism is \cite{Tim1}
\be
\label{GSds2}
d s^2 = -d t^2 + \frac{a_{\parallel}^2(t,r)}{(1-\kappa r^2)} d r^2 + a_{\perp}^2(t,r)
r^2 d \Omega^2 .
\ee
This is the same metric in the same coordinates as that used in this paper \eq{LTBInterval}. This allows us to compare directly the perturbed metric components once the relations between the background functions are known. In the rest of this section, where a symbol is used in the Clarkson, Clifton and February formalism which has a different meaning to the same symbol in this paper we have made it calligraphic, except for ``$v$'' which is made ``$\rm{v}$''. Also, a radial derivative is later defined which differs slightly from that used in earlier sections of this paper. To distinguish this alternative radial derivative we use a dagger in place of the prime used in Ref.~\cite{Tim1}. From \eq{GSds2} and \eq{LTBInterval} we get,

\begin{equation}
\label{sfconvert}
\sfx = \frac{a_{\parallel}}{\sqrt{(1-\kappa r^2)}} \,, \qquad
\sfy = a_{\perp} r \,, \qquad
H_\parallel \equiv \frac{\dot{a_{\parallel}}}{a_{\parallel}} = H_\sfx \,, \qquad
H_\perp \equiv \frac{\dot{a_{\perp}}}{a_{\perp}} = H_\sfy \,,
\end{equation}
where $\kappa \equiv \kappa (r)$. The radial derivative defined in Ref.~\cite{Tim1} for an arbitrary function, $F$, is
\be
\label{RadDer}
F^\dagger = \frac{\sqrt{(1-\kappa r^2)}}{a_{\parallel}} F' = \frac{F '}{\sfx} \,, 
\ee
where the time derivative of the above radial derivative behaves as
\be
\label{RadDerDot}
(\dot{F})^\dagger - (F^\dagger \dot{)} = H_\parallel F^\dagger = H_\sfx \frac{F '}{\sfx} \,,
\ee

\subsection{Perturbations}
\label{GSPert}

The perturbed portion of the metric for axial perturbations is given as
\begin{align}
\label{GSpertmetax}
\delta g_{\mu \nu} 
&\equiv
\left( \begin{array}{cc}
0 & h_A^{\text{axial}} \bar {\cal{Y}}_a\\ \nonumber
h_A^{\text{axial}} \bar {\cal{Y}}_a \; & h \; \bar {\cal{Y}}_{ab} \end{array} \right)\\
\end{align}
and for the polar perturbations as
\begin{align}
\label{GSpertmetpol}
\delta g_{\mu \nu} 
&\equiv
\left( \begin{array}{cc}
h_{AB} {\cal{Y}} & h_A^{\text{polar}} {\cal{Y}}_{a}\\ \nonumber
h_A^{\text{polar}} {\cal{Y}}_{a} \; \; & a_{\perp}^2 r^2 (K {\cal{Y}} \gamma_{ab}+G {\cal{Y}}_{:ab}) \end{array} \right)\\
\end{align}
In the above equations ${\cal{Y}} \equiv {\cal{Y}}^{(lm)}$ and are the various spherical harmonic functions for scalar vector and tensor perturbations (see Ref.~\cite{Tim1}). The index $A$ runs over $t$ and $r$, while $a$ runs over $\theta$ and $\phi$. The colon represents the covariant derivative with respect to the metric on the unit sphere. $h_A^{\text{axial}} , h , h_{AB} , h_A^{\text{polar}} , K , G $ are the perturbation variables and are functions of $x^A$.
By direct comparison between the perturbed metrics in both formalisms we find,
\be
\label{psiGS}
\psi = \frac{1}{3} (C_{rr} + C_{\theta \theta} + C_{\phi \phi}) = \frac{1}{6} \left(\frac{h_{r r} {\cal{Y}}}{\sfx^2} + \frac{h \; \bar {\cal{Y}}_{\theta \theta}}{\sfy^2}+ \frac{h \; \bar {\cal{Y}}_{\phi \phi}}{\sfy^2 \sin^2 \theta} + 2 K {\cal{Y}} + G {\cal{Y}}_{:\theta \theta} + \frac{G {\cal{Y}}_{:\phi \phi}}{\sin^2 \theta} \right)
 ,
\ee
where we have used Bondi's scale factors, $\sfx$ and $\sfy$, for brevity. The covariant form of the axial perturbed 4-velocities is
\be
\label{4vaxGMGcov}
\delta u_{\mu} = (0, {\bar{\rm{v}}} \; \bar {\cal{Y}}_a)
\ee
and the scalar perturbed 4-velocities are
\be
\label{4vscalGMGcov}
\delta u_{\mu} = \left[ \left(\tilde{w} {\hat{n}}_A+\frac{1}{2}h_{AB} {\hat{u}}^B
\right) {\cal{Y}}, \tilde{\rm{v}} \; {\cal{Y}}_a \right] ,
\ee
where ${\bar{\rm{v}}}, \tilde{w}, \tilde{\rm{v}}$ are all functions of $x^A$, and $\hat{n}_A$ is the unit spacelike radial vector and $\hat{u}^A$ is the unit timelike vector.
The contravariant form of the perturbed metric for axial perturbations is
\begin{align}
\label{GSpertmetaxuppy}
\delta g^{\mu \nu} 
&\equiv
\left( \begin{array}{cccc}
0 & 0 & \frac{1}{\sfy^2} h_t^{\text{axial}} \bar {\cal{Y}}_\theta & \frac{1}{\sfy^2 \sin^2 \theta} h_t^{\text{axial}} \bar {\cal{Y}}_\phi\\ \nonumber
0 & 0 & -\frac{1}{\sfx^2 \sfy^2} h_r^{\text{axial}} \bar {\cal{Y}}_\theta & -\frac{1}{\sfx^2 \sfy^2 \sin^2 \theta} h_r^{\text{axial}} \bar {\cal{Y}}_\phi\\ \nonumber
\frac{1}{\sfy^2} h_t^{\text{axial}} \bar {\cal{Y}}_\theta & -\frac{1}{\sfx^2 \sfy^2} h_r^{\text{axial}} \bar {\cal{Y}}_\theta & - \frac{1}{\sfy^4} h \; \bar {\cal{Y}}_{\theta \theta} \; & - \frac{1}{\sfy^4 \sin^2 \theta} h \; \bar {\cal{Y}}_{\theta \phi}\\ \nonumber
\frac{1}{\sfy^2 \sin^2 \theta} h_t^{\text{axial}} \bar {\cal{Y}}_\phi & -\frac{1}{\sfx^2 \sfy^2 \sin^2 \theta} h_r^{\text{axial}} \bar {\cal{Y}}_\phi & - \frac{1}{\sfy^4 \sin^2 \theta} h \; \bar {\cal{Y}}_{\theta \phi} & - \frac{1}{\sfy^4 \sin^4 \theta} h \; \bar {\cal{Y}}_{\phi \phi} \end{array} \right)\\
\end{align}
and for the polar perturbations is
\begin{align}
\label{GSpertmetpoluppy}
\delta g^{\mu \nu} 
&\equiv
\left( \begin{array}{cccc}
- h_{tt} {\cal{Y}} & \frac{1}{\sfx^2} h_{tr} {\cal{Y}} & \frac{1}{\sfy^2} h_t^{\text{polar}} {\cal{Y}}_{\theta} \; \; & \frac{1}{\sfy^2 \sin^2 \theta} h_t^{\text{polar}} {\cal{Y}}_{\phi}\\ \nonumber
\frac{1}{\sfx^2} h_{tr} {\cal{Y}} & - \frac{1}{\sfx^4} h_{rr} {\cal{Y}} & - \frac{1}{\sfx^2 \sfy^2} h_r^{\text{polar}} {\cal{Y}}_{\theta} \; \; & - \frac{1}{\sfx^2 \sfy^2 \sin^2 \theta} h_r^{\text{polar}} {\cal{Y}}_{\phi}\\ \nonumber
\frac{1}{\sfy^2} h_t^{\text{polar}} {\cal{Y}}_{\theta} \; \; & - \frac{1}{\sfx^2 \sfy^2} h_r^{\text{polar}} {\cal{Y}}_{\theta} \; \; & - \frac{1}{\sfy^2} (K {\cal{Y}} +G {\cal{Y}}_{:\theta \theta}) \; \; & - \frac{1}{\sfy^2 \sin^2 \theta} G {\cal{Y}}_{: \theta \phi}\\ 
\nonumber
\frac{1}{\sfy^2 \sin^2 \theta} h_t^{\text{polar}} {\cal{Y}}_{\phi} & - \frac{1}{\sfx^2 \sfy^2 \sin^2 \theta} h_r^{\text{polar}} {\cal{Y}}_{\phi} & - \frac{1}{\sfy^2 \sin^2 \theta} G {\cal{Y}}_{: \theta \phi} & - \frac{1}{\sfy^2 \sin^4 \theta} (K {\cal{Y}} \sin^2 \theta +G {\cal{Y}}_{: \phi \phi}) \end{array} \right)\\
\end{align}
where we have once again used Bondi's scale factors, $\sfx$ and $\sfy$, for brevity.
The perturbed 4-velocity in contravariant form is
\bea
\label{4vGMGcont}
u^{\mu} &=& \Bigg[ \qquad 1 + \frac{1}{2} h_{t t} {\cal{Y}}, \qquad \qquad \qquad \qquad \qquad \qquad - \frac{{\cal{Y}}}{\sfx^2} \left(\frac{1}{2} h_{t r} + \sfx \tilde{w} \right) ,\\ \nonumber & & \frac{1}{\sfy^2} \left( {\bar{\rm{v}}} \; \bar {\cal{Y}}_\theta + \tilde{\rm{v}} \; {\cal{Y}}_\theta -  h_t^{\text{axial}} \bar {\cal{Y}}_\theta - h_t^{\text{polar}} {\cal{Y}}_{\theta}  \right), \qquad \frac{1}{\sfy^2 \sin^2 \theta} \left( {\bar{\rm{v}}} \; \bar {\cal{Y}}_\phi + \tilde{\rm{v}} \; {\cal{Y}}_\phi -  h_t^{\text{axial}} \bar {\cal{Y}}_\phi - h_t^{\text{polar}} {\cal{Y}}_{\phi}  \right)  \Bigg] ,
\eea
where the last three terms correspond directly with $v^r, v^\theta, v^\phi$ respectively in the formalism of this paper.
Substituting \eq{4vGMGcont}, \eq{psiGS}, \eq{RadDer} and \eq{sfconvert} into \eq{zetafull4} we get
\bea
\label{GSzetafull}
- \SMTP &=& \frac{1}{6} \left( \frac{(1-\kappa r^2)}{{a^2}_{\parallel}} h_{r r} {\cal{Y}} + \frac{h \; \bar {\cal{Y}}_{\theta \theta}}{{a_{\perp}}^2 r^2}+ \frac{h \; \bar {\cal{Y}}_{\phi \phi}}{{a_{\perp}}^2 r^2 \sin^2 \theta} + 2 K {\cal{Y}} + G {\cal{Y}}_{:\theta \theta} + \frac{G {\cal{Y}}_{:\phi \phi}}{\sin^2 \theta} \right) + \frac{\delta \rho}{3\bar{\rho}} \\ 
\nonumber
&+& \frac{1}{3}\Bigg\{ \partial_{\theta} \int \frac{1}{{a_{\perp}}^2 r^2} \left( {\bar{\rm{v}}} \; \bar {\cal{Y}}_\theta + \tilde{\rm{v}} \; {\cal{Y}}_\theta -  h_t^{\text{axial}} \bar {\cal{Y}}_\theta - h_t^{\text{polar}} {\cal{Y}}_{\theta}  \right) dt \\ \nonumber &+& \partial_{\phi} \int \frac{1}{{a_{\perp}}^2 r^2 \sin^2 \theta} \left( {\bar{\rm{v}}} \; \bar {\cal{Y}}_\phi + \tilde{\rm{v}} \; {\cal{Y}}_\phi -  h_t^{\text{axial}} \bar {\cal{Y}}_\phi - h_t^{\text{polar}} {\cal{Y}}_{\phi}  \right) dt \\ 
\nonumber &+& \cot \theta \int \frac{1}{{a_{\perp}}^2 r^2} \left( {\bar{\rm{v}}} \; \bar {\cal{Y}}_\theta + \tilde{\rm{v}} \; {\cal{Y}}_\theta -  h_t^{\text{axial}} \bar {\cal{Y}}_\theta - h_t^{\text{polar}} {\cal{Y}}_{\theta}  \right) dt - \partial_r \int \frac{{\cal{Y}}(1-\kappa r^2)}{{a_{\parallel}}^2} \left(\frac{1}{2} h_{t r} + \frac{a_{\parallel}}{\sqrt{(1-\kappa r^2)}} \tilde{w} \right) dt \\ 
\nonumber &-& \left(\left(\frac{a_{\parallel}}{\sqrt{(1-\kappa r^2)}}\right)^\dagger + 2 \frac{(a_{\perp} r)^\dagger a_{\parallel}}{a_{\perp} r \sqrt{(1-\kappa r^2)}}
+ \frac{{\bar{\rho}}^\dagger a_{\parallel} }{\bar{\rho} \sqrt{(1-\kappa r^2)}} \right) \int \frac{(1-\kappa r^2)}{{a^2}_{\parallel}} {\cal{Y}} \left(\frac{1}{2} h_{t r} + \frac{a_{\parallel}}{\sqrt{(1-\kappa r^2)}} \tilde{w} \right) dt \Bigg\} \,,
\eea
which is our gauge invariant quantity, conserved on all scales with only adiabatic pressure perturbations, but expressed in terms of the perturbation functions used in \cite{Tim1}. As mentioned in \ref{LTBst}, this relates directly to the density perturbation on constant curvature hypersurfaces through \eq{GIdenspertZeta2}
\begin{equation*}
\delta \tilde{\rho} \Big|_{\psi=0} = - 3 \bar{\rho} \SMTP\, .
\end{equation*}
\eq{GSzetafull} is clearly more complicated than \eq{zetafull4}.






\end{document}